\documentclass[aps,prl,reprint,twocolumn,showpacs,floatfix,amsmath,amssymb,superscriptaddress,showkeys,groupedaddress]{revtex4-1}
\usepackage{graphicx}
\usepackage{epstopdf}
\usepackage{color}
\usepackage{hyperref}
\usepackage{bm}
\usepackage{printlen}
\usepackage{units}
\usepackage{bbm,pifont}
\usepackage{tikz}

\begin{document}
\title{Phonon Magnetic Moment from Electronic Topological Magnetization}

\author{Yafei Ren}
\affiliation{Department of Physics, The University of Texas at Austin, Austin, Texas 78712, USA}

\author{Cong Xiao}
\affiliation{Department of Physics, The University of Texas at Austin, Austin, Texas 78712, USA}

\author{Daniyar Saparov}
\affiliation{Department of Physics, The University of Texas at Austin, Austin, Texas 78712, USA}

\author{Qian Niu}
\affiliation{Department of Physics, The University of Texas at Austin, Austin, Texas 78712, USA}

\date{\today}

\begin{abstract}
The traditional theory of magnetic moments for chiral phonons is based on the picture of the circular motion of the Born effective charge, typically yielding a small fractional value of the nuclear magneton. Here we investigate the adiabatic evolution of electronic states induced by lattice vibration of a chiral phonon and obtain an electronic orbital magnetization in the form of a topological second Chern form.  We find that the traditional theory needs to be refined by introducing a $\bm{k}$ resolved Born effective charge, and identify another contribution from the phonon-modified electronic energy together with the momentum-space Berry curvature.
The second Chern form can diverge when there is a Yang's monopole near the parameter space of interest as illustrated by considering a phonon at the Brillouin zone corner in a gaped graphene model. 
{ We also find large magnetic moments for the optical phonon in bulk topological materials where non-topological contribution is also important. The magnetic moment experiences a sign change when the band inversion happens.}
\end{abstract}

\maketitle

Lattice phonons are commonly known to carry a well-defined crystal momentum and energy quantum, and can couple to lights through a time-varying electrical dipole moment described by the Born effective charge $Q^*$~\cite{BornHuang_Book}. 
Recently, phonon chirality has attracted much attention both {theoretically~\cite{Phonon_AngularMom_Lifa_14, Phonon_Chiral_AngularMom_Lifa_15, ChiralPhonon_Kekule_17, ChiralPhonon_r3xr3_18, ChiralPhonon_GhBN_18, ChiralPhonon_Kagome_19, Kagome_ChiralPhonon_21, DanPhonon} and experimentally~\cite{ChiralPhonon_Exp_18, Phonon-Magnon_Exp_18, ChiralPhonon_ExcitonReplicaTMD_19, Phonon_MagneticField_17, Phonon_surface_21, PHE_Exp_05, PHE_Exp_07,PHE_SpinLiquid_Exp_17, THE_Cuprate_20}.
It can interact with the electronic valley degree of freedom and affect valley excitons~\cite{ChiralPhonon_Exp_18, ChiralPhonon_ExcitonReplicaTMD_19}.
It can also couple strongly with electron spins and can be employed to control magnetism in magnetic materials~\cite{Phonon_MagneticField_17, Phonon_surface_21}. 
In particular, the experiments reveal the chirality of phonon under a magnetic field through thermal Hall effect in, e.g., the pseudogap phase of cuprates~\cite{PHE_Exp_05, PHE_Exp_07, PHE_SpinLiquid_Exp_17, THE_Cuprate_20}. 

One natural way to characterize the coupling of phonon to a magnetic field is through the phonon magnetic moment, defined for example by the phonon energy shift under a magnetic field~\cite{PhononMagnetization_20, PhononMagPbTe}.}
For a phonon with nonzero angular momentum $L$, one would expect a phonon magnetic moment in the order of ionic magneton $\frac{Q^*L}{m_{\rm I}}$ where $Q^*$ and $L$ are generally in the order of electron charge~\cite{Phonon_OrbitMoment_19} and $\hbar$~\cite{Phonon_AngularMom_Lifa_14, Kagome_ChiralPhonon_21}, respectively. The ion mass $m_{\rm I}$ is much larger than the electron mass~\cite{PhononBfield_Multiferroicity_17, Phonon_OrbitMoment_19, PhononBfield_4fPM_20, Cong_20}. However, recent experiments suggest that the phonon magnetic moment can be three to four orders of magnitude larger~\cite{PhononMagnetization_20, PhononMagPbTe}, which calls for a deeper understanding of this physical concept.

In this Letter, we formulate the phonon magnetic moment as electronic magnetization in an adiabatic response to the underlying ionic circular motion, focusing on the orbital part. We find that the traditional theory needs to be refined in terms of a momentum resolved Born effective charge, and recognize an extra contribution due to phonon-induced electron energy coupled to the electronic Berry curvature in momentum space. These contributions are captured by a topological second Chern form,  which can be very large when there is a Yang's monopole near the parameter space of interest as demonstrated by studying the phonon at the Brillouin zone corner in a gaped graphene model, where only the newly identified contribution is nonzero. {We also find a large magnetic moment for the optical phonon in topological materials where non-topological electronic contributions are also important. 
}

\textit{\textbf{Adiabatic current pumping by phonon.---}} The phonon magnetic moment refers to the variation of the total magnetic moment when a phonon is created, which can be contributed by the circular motion of the ions, phonon pumped electronic magnetization from spin~\cite{Phonon_Spin_20} and orbital effect~\cite{PhononBfield_Multiferroicity_17, Phonon_OrbitMoment_19, PhononBfield_4fPM_20, Cong_20, PhononBfield_Multiferroicity_17, Phonon_OrbitMoment_19, PhononMag_Vanderbilt_18, Mag_Inhomogeneity_Di_09, Dong_Niu_18, OrbMag_Adiabatic_19}. 
The orbital contribution can be separated into a non-topological and a topological part~\cite{OrbMag_Adiabatic_19}.  
The former shows a similar form as that from spin~\cite{Cong_20, Phonon_Spin_20, NoteSpinOrb}.
{The latter however involves gauge-dependent Berry connection~\cite{OrbMag_Adiabatic_19}. 
Here, we focus on the latter and provide an explicitly gauge-independent form of the topological magnetization induced by phonon.}

To have nonzero out-of-plane orbital magnetization, the time-reversal invariance and the mirror symmetry about any perpendicular mirror plane need to be broken in the presence of a phonon. Phonons with chirality typically satisfy the criteria.
We consider a phonon mode with a known polarization vector. The ions' motion is parameterized by $\bm{u}=(u_x(t,\bm{r}), u_y(t, \bm{r}))$ where $u_{x,y}$ can be the displacement of one representative atom that is periodic temporally. 
{We assume $u_{x,y}$ to be slowly varying spatially in the following derivations and take it to be uniform (e.g., optical phonon near the $\Gamma$ point) in the final expression of the phonon magnetic moment.
When the electronic band gap is larger than the phonon energy, the electronic state evolves adiabatically following the ion governed by the Hamiltonian $H(\bm{k},\bm{u})$ with $\bm{k}$ being the momentum.  }

We define the magnetization $\bm{M}$ by employing the constituent equation $\bm{j}=\partial_t \bm{P} + \nabla \times \bm{M}$ where $\bm{P}$ is polarization and $\bm{j}$ is bounded current density. 
By employing the semiclassical theory of Bloch electrons, the topological local current following Ref.~\onlinecite{Mag_Inhomogeneity_Di_09} can be expressed as
\begin{align} \label{j2B}
j_\alpha^{(2)}&=\sum_{\delta} e\dot{u}_\delta\int \frac{d \bm{k}}{(2\pi)^2} \Omega_{k_\alpha k_\beta r_{\beta} u_\delta}
\end{align} 
where $e$ is the elementary charge with a positive sign, $k_{\alpha}$, $r_{\alpha}$, and $u_{\alpha}$ are the momentum, real space coordinate, and the displacement along the $\alpha$-th direction.
$\dot{u}_\delta$ represents the time derivative of $u_\delta$ with $\delta$ being $(x,y)$.
By writing the subscripts in a general form for simplicity, $\Omega_{\alpha \beta {\gamma} \delta}= \Omega_{\alpha \beta} \Omega_{\gamma \delta} + \Omega_{\beta\gamma} \Omega_{\alpha \delta} - \Omega_{\alpha \gamma} \Omega_{\beta\delta}$.
$\Omega_{\alpha\beta}=\partial_\alpha A_\beta - \partial_\beta A_\alpha$ is the abelian Berry curvature of the corresponding indices where $A_\alpha=\langle \varphi_{\bm{k}}| i\partial_\alpha |\varphi_{\bm{k}} \rangle$ is the Berry connection with $|\varphi_{\bm{k}} \rangle$ being the periodic part of the Bloch wavefunction. 

\textit{\textbf{Phonon magnetic moment.---}}
In Eq.~\ref{j2B}, the Berry curvatures are evaluated at each $\bm{k}$ with finite $\bm{u}$. As the displacement $\bm{u}$ is extremely small compared to the lattice constant, we thus can perform a Taylor expansion at $\bm{u}=0$, which is reasonable as long as $\bm{u}$ does not close the band gap. 
To the first order of the expansion, the current reads
\begin{align} \label{j2expansion}
 j_\alpha^{(2)}&= \sum_{\delta} e \dot{u}_\delta \int \frac{d\bm{k}}{(2\pi)^2} \Omega_{k_\alpha k_\beta r_{\beta} u_\delta}|_{\bm{u}=0} \nonumber \\
 &+ \sum_{\delta \gamma} e\dot{u}_\delta u_\gamma \int \frac{d\bm{k}}{(2\pi)^2} \partial_{u_\gamma} \Omega_{k_\alpha k_\beta r_{\beta} u_\delta}|_{\bm{u}=0}
\end{align} 
where the Berry curvatures are evaluated at $\bm{u}=0$ that is the case for all the Berry curvatures hereafter. 
The first line is a time derivative term that corresponds to the current density from electrical polarization.
In the second line, by symmetrizing the summation with respect to $(\delta,\gamma)$, one can obtain a second-harmonic polarization current density that is symmetric about exchanging $(\delta,\gamma)$ 
and a magnetization current density that is anti-symmetric. 
The latter gives rise to the time-averaged out-of-plane magnetization~\cite{SM}
\begin{align} \label{Mz}
M_{z} &= \frac{e}{2m_{\rm I}} L_{\rm I} \int \frac{d\bm{k}}{(2\pi)^2} \Omega_{k_\alpha k_{\beta}  u_x u_y}
\end{align}
where $m_{\rm I}$ is the mass of the representative ion with averaged angular momentum ${L_{\rm I}}=\frac{m_{\rm I}}{T}\int_0^T (\bm{u}\times \dot{\bm{u}})_z dt$ over the phonon period $T$. 
The integral of $M_z$ over the sample size gives rise to the phonon magnetic moment. 

Equation~\ref{Mz} indicates that the linearly polarized phonon with zero angular momentum shows zero magnetic moments.
The gauge invariant second Chern form $\Omega_{k_\alpha k_{\beta}  u_x u_y}$ is evaluated at $\bm{u}=0$ that is thus an intrinsic property of the electronic system. In contrast to $\Omega_{k_x k_y}$, time reversal symmetry guarantees that $\Omega_{k_\alpha k_{\beta}  u_x u_y}(\bm{k})=\Omega_{k_\alpha k_{\beta}  u_x u_y}(-\bm{k})$. Thus, phonons in nonmagnetic system can also have magnetic moment. 

\begin{figure}
    \centering
    \includegraphics[width=8 cm]{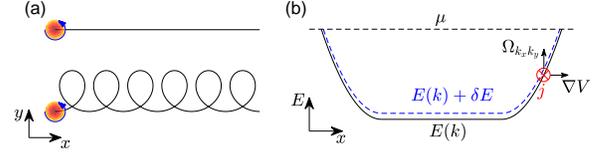}
    \caption{Physical picture of the phonon magnetic moment. (a) With a phonon, the trajectory of the center of mass of a wavepacket (straight line on top panel) is superposed by a circular orbit in the lower panel. (b) By modifying the electronic energy, phonon changes the boundary confinement potential $V$ induced current in the presence of momentum-space Berry curvature.} 
    \label{Orbit}
\end{figure}

Here we show an intuitive understanding of the phonon magnetic moment. The second Chern form reads explicitly $\Omega_{k_xk_yu_xu_y} = \Omega_{k_x u_y} \Omega_{k_y u_x} - \Omega_{k_x u_x} \Omega_{k_y u_y}  + \Omega_{k_x k_y} \Omega_{u_x u_y}$. The first two terms depend only on $\Omega_{k_i u_j}$, whose average gives rise to the macroscopic Born effective charge tensor $\bm{Q}^*$ with element $Q_{ij}^*=e\int\frac{d\bm{k}}{(2\pi)^2}{\Omega}_{k_i u_j}$~\cite{BornEffCharge_DFT_97, BornEffCharge_BerryCurv_02, Phonon_ElectronCoupling_BornCharge_19, Phonon_Polarization_20} that is related to the macroscopic polarization $\bm{P}=\bm{Q}^*\bm{u}$.
The electric dipole moment contributed from each wavepacket is thus $e \bm{\Omega} \bm{u}$. Therefore, we identify $e \bm{\Omega}$ as the $\bm{k}$-resolved Born effective charge tensor with matrix element $e\Omega_{k_i u_j}$.
Such a dipole moment suggests that the mass center of a wavepacket deflects its trajectory by $\bm{d}=-\bm{\Omega} \bm{u}$, which form a circular orbit as illustrated in Fig.~\ref{Orbit}(a). The corresponding orbital magnetic moment from this orbit is $\frac{-e}{2}(\bm{d}\times \dot{\bm{d}})_z$ that equals to the first two terms of $\Omega_{k_xk_yu_xu_y}$.
{In an atomic crystal, this term cancels the magnetic moment from charged ion~\cite{SM}.}
It is noteworthy that, near the gap closing points the Berry curvature $\Omega_{k_iu_j}$ can be large. In this case, although the integration of $\Omega_{k_iu_j}$, i.e., $\bm{Q}^*$, is usually in the order of ionic charge, the integral of $\Omega_{k_iu_j}\Omega_{k_ju_i}$ can be extremely large, which is different from the phonon magnetic estimated by $\bm{Q}^*$~\cite{PhononBfield_Multiferroicity_17, Phonon_OrbitMoment_19}. 

The contribution shown above can find its position in the modern theory of the orbital magnetization $M$ of a two-dimensional system~\cite{ModernTheory_Di_05, ModernTheory_Resta_05,ModernTheory_10,Xiao_RMP}. At zero temperature, 
\begin{align} \label{Mag}
M=\int\frac{d\bm{k}}{(2\pi)^2} [m(\bm{k})+\frac{e}{\hbar} (\mu-E(\bm{k}))\Omega_{k_x k_y}] 
\end{align}
where $E(\bm{k}) $ identifies the energy bands below the chemical potential $\mu$ and $m(\bm{k})$ is the orbital magnetic moment from the self-rotation of each wavepacket. 
Our results suggest that the $m(\bm{k})$ term should be refined to include the magnetic moment from the orbital motion of the center of mass of each wavepacket.

The second term in the magnetization $M$ is topological that can be interpreted as the boundary current contribution in the presence of boundary confinement potential $V$ and  nonzero $\Omega_{k_x k_y}$ as illustrated in Fig.~\ref{Orbit}(b). Phonon can also carry magnetic moment from the boundary current by modifying the electronic energy through the geometrical phase. In a period of $\bm{u}$, the electronic state will pick up a phase factor $e^{-iE(\bm{k})T/\hbar+i\eta}$ composed of the dynamical phase and the geometrical one $\eta=\Omega_{u_x u_y} S_u$ where $S_u=\frac{1}{2} \int_0^T(\bm{u}\times \dot{\bm{u}})_z dt$ represents the area swept by $\bm{u}$ in a period. The total phase can be regarded as the dynamical phase from a modified energy $E+\delta E$ with the energy correction being {$\delta E=-\hbar \eta/T=-\hbar\Omega_{u_xu_y}\frac{L_{\rm I}}{2m_{\rm I}}$. By changing the energy $E(k)$ in Eq.~\eqref{Mag} to the corrected one, i.e., $E+\delta E$, one can obtain the term proportional to $\Omega_{k_xk_y}\Omega_{u_xu_y}$.}

{
\textbf{\textit{Nonabelian formulas.---}}
The above discussions are restricted to the case of a single occupied band. When multi-bands are occupied, the topological and nontopological contributions from each band should be regrouped to enforce $U(N)$ gauge invariance within the occupied $N$-dimensional Hilbert space~\cite{DiXiao_20}.
As a result, the topological contribution becomes the nonabelian one 
\begin{align} 
M_{z} &= \frac{e}{2m_{\rm I}} L_{\rm I} \int \frac{d\bm{k}}{(2\pi)^2} ~ {\rm Tr} \, \Omega_{k_\alpha k_{\beta}  u_x u_y}
\end{align}
with nonabelian Berry curvatures $\Omega_{\alpha\beta}=\partial_\alpha A_\beta-\partial_\beta A_\alpha - i[A_\alpha, A_\beta]$. The Berry connection $A_\alpha$ is a matrix, $A_\alpha^{mn}=\langle \varphi_m | i\partial_\alpha |\varphi_n\rangle$, with $(m,n)$ being the indices of the occupied bands.
The non-topological contribution for single-occupied band case~\cite{Cong_20} should be generalized to
\begin{align}
    M_z^{nt} &=-\frac{e}{2m_{\rm I}} L_{\rm I} (\partial_{u_x}F_{u_y}-\partial_{u_y}F_{u_x}) \\
    F_{u_i} &={\rm Re}\sum_{n\in {\rm occu}} \sum_{n',m' \in {\rm unoc}} \int \frac{d\bm{k}}{(2\pi)^2} \nonumber \\
    &\times \frac{\langle n| \partial_{u_i}H |n' \rangle [ (\bm{v}_{n'm'} + \bm{v}_{nn} \delta_{n'm'}) \times \bm{v}_{m'n}]_z } {(E_n-E_{n'})^2(E_n-E_{m'})} \nonumber
\end{align}
where $\bm{v}_{mn}= \langle m | \nabla_k H | n \rangle$ is a matrix element of the velocity operator and $\rm{Re}$ means the real part. 
These results are consistent with the theory in Ref.~\onlinecite{OrbMag_Adiabatic_19}. One can see this by applying the latter to the phonon, expanding to the first order of $u_{x,y}$, and taking the anti-symmetric part (the symmetric part vanishes under time average)~\cite{Note, OME_10}.
Comparing with the formulas in Ref.~\onlinecite{OrbMag_Adiabatic_19}, the present results are explicitly gauge-invariant and are easier to be adopted by first-principles calculations.
}

\begin{figure}
    \centering
    \includegraphics[width=8.5 cm]{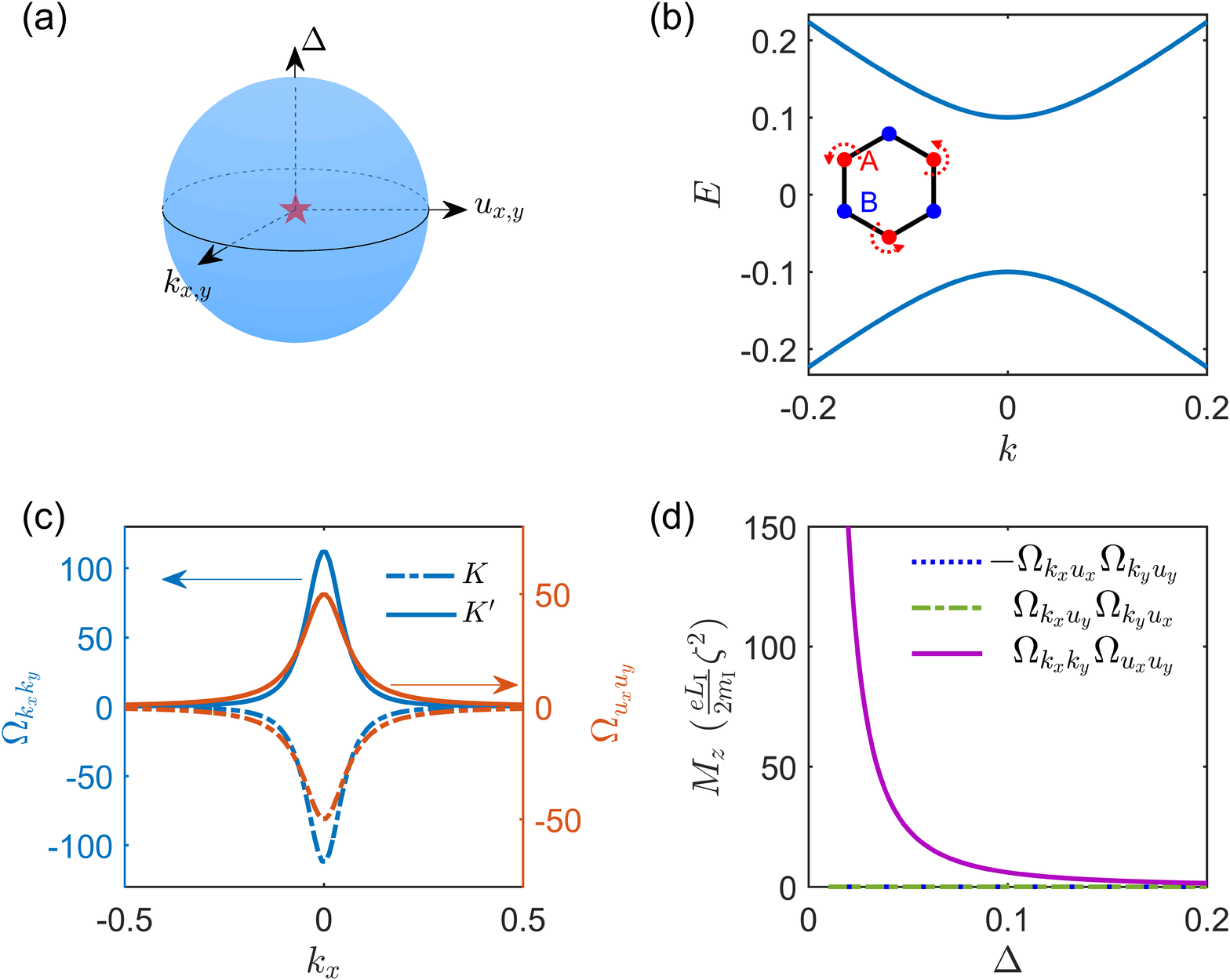}
    \caption{(a) A sphere enclosing a Yang's monopole (red pentagram) in the five-dimensional parameter space formed by $(\Delta, k_x, k_y, u_x, u_y)$. (b) Energy band of a gapped graphene with $K/K'$ valleys being folded to the zone center. Inset shows the polarization vector of the chiral phonon in the Brillouin zone corner where B atoms stay still. (c) Berry curvature $\Omega_{k_x k_y}$ and $\Omega_{u_x u_y}$ along $k_x$. (d) Contributions to the magnetization from different terms in the second Chern form.}
    \label{chiralPhonon}
\end{figure}

\textit{\textbf{Divergence near Yang's monopole.---}} 
The topological nature of the second Chern form allows the presence of a large phonon magnetic moment. By integrating the second Chern form over a four-sphere around a Yang's monopole, one can obtain an integer~\cite{YangMonopole_18}. The second Chern form can thus become divergently large close to the monopole similar to the Berry curvature near a Weyl point~\cite{SemimetalBC_16}. Near the monopole, the effective Hamiltonian reads $H=\textbf{q}\cdot \bf{\Gamma}$ where $\bm{\Gamma}$ are Dirac matrices with $\Gamma_{1\sim 5}=(\sigma_x \tau_z, \sigma_y \tau_z, \sigma_z \tau_z, \sigma_0 \tau_x, \sigma_0 \tau_y)$, $\bm{\sigma}$ and $\bm{\tau}$ being Pauli matrices. 
By taking $\bm{q}=(v_F k_x, v_F k_y, \Delta, \zeta u_y, -\zeta u_x)$, this Hamiltonian of $H$ can be mapped to the effective model of graphene with chiral phonon at the Brillouin zone corner 
\begin{align} \label{Heff}
    H_{\rm eff} =\bm{q}\cdot \bm{\Gamma} =\left[ \begin{array}{cccc} 
             \Delta & v_F \pi^\dagger & \zeta \rho^\dagger & \\
             v_F \pi & -\Delta & & \zeta \rho^\dagger \\
             \zeta \rho & & - \Delta & -v_F \pi^\dagger \\
              & \zeta \rho & - v_F \pi & \Delta 
             \end{array} \right] 
\end{align}
under the basis $\{|K,B\rangle, |K,A\rangle, |K',A\rangle, |K',B\rangle \}$ in spin up sector. Here, $v_F=-3t_0/2$ is the Fermi velocity, $\Delta$ stands for the sublattice potential, and $\pi=k_x+ik_y$. The chiral phonon leads to the intervalley coupling with $\rho=u_y+iu_x$ and $\zeta=-3t_0\lambda/2$. Here, $\bm{u}$ is the displacement of the A atom at the top left corner as shown in the inset of Fig.~\ref{chiralPhonon}(b) and the displacements of the other atoms are expressed as functions of $u_{x,y}$. Due to finite momentum of the $K$-valley chiral phonon, the neighboring A atoms show displacements with phase differences of $e^{\pm2i\pi/3}$ forming a $\sqrt{3}\times \sqrt{3}$ superlattice at nonzero $\bm{u}$. Thus, $K/K'$ valleys of graphene electronic bands are folded to the zone center~\cite{RenPRB15, ZhouShuyun1}. The energy bands are shown in Fig.~\ref{chiralPhonon}(b) where the bands are doubly degenerate and the valley is still a good quantum number in the $\bm{u}=0$ limit.

{The non-topological contribution to the phonon magnetic moment vanishes. In the topological contribution, the wavepacket Born effective charge $e\Omega_{k_i u_j}(\bm{k})$ vanishes whereas the boundary current part is large.}
As shown in Fig.~\ref{chiralPhonon}(c), $\Omega_{k_x k_y}$ are nonzero with opposite signs for opposite valleys. 
Meanwhile, the Berry curvature $\Omega_{u_x u_y}$ is also nonzero and valley polarized. The phonon magnetic moment is proportional to {$\rm{sign}(\Delta)\frac{1}{12\pi} \frac{\zeta^2}{\Delta^2}$ with $\rm{sign}(\Delta)=\pm 1$ being the sign of the mass term}. 
The magnetic moment thus diverges as $\Delta$ goes to zero as plotted in Fig.~\ref{chiralPhonon}(d) with $\lambda=1$. 

It is noted that, as the adiabatic approximation is employed, our results break down as the band gap becomes smaller than the phonon energy. Specific to graphene, the chiral phonon energies range from $100\sim200~$meV, which corresponds to a $\Delta=0.02\sim 0.04~t_0$ with $t_0=2.6~$eV. Our results in Fig.~\ref{chiralPhonon}(d) is shown down to the lower limit.
By considering the other spin sector, the phonon magnetic moment doubles. When a more realistic $\lambda=3$~\cite{Graphene_Strain_Model_09} for graphene is employed, i.e., $\zeta \simeq 12~$eV/\AA, the result increases further by one order. Thus, the magnetic moment for a chiral phonon in graphene can reach $10^{3}$ times larger than the atomic magneton, which is in the (sub)order of the electronic magneton.

{
\begin{figure}
    \centering
    \includegraphics[width=8.5 cm]{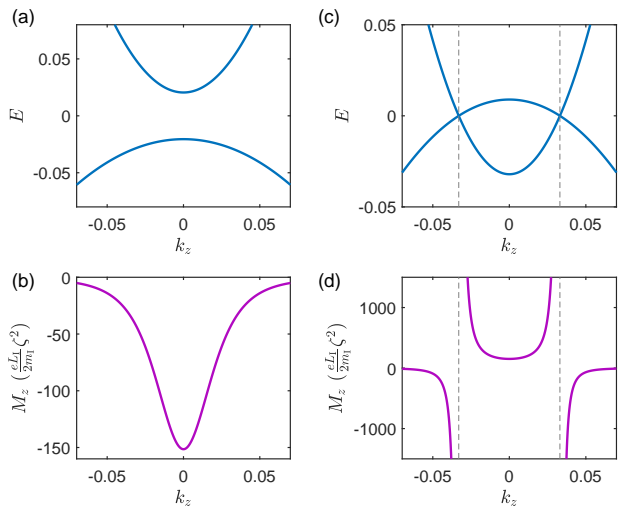}
    \caption{(a) and (b) Electronic structure and contribution to the phonon magnetic moment along $\Gamma$-$Z$ direction with $\Delta_0=-0.0205~$eV. (c) and (d) Electronic structure and contribution to the phonon magnetic moment along $\Gamma$-$Z$ direction for Cd$_3$As$_2$ with $\Delta_0=0.0205~$eV. Dashed vertical lines indicate the position of the Dirac points.
    In the calculation, $\Delta_1=18.77~$eV\AA$^2$, $\epsilon_0=-0.0116~$eV, $\epsilon_1=10.59~$eV\AA$^2$, and $v_F=0.889~$eV\AA. 
    }
    \label{fig:Cd3As2V1}
\end{figure}

\textit{\textbf{Phonon magnetic moment in bulk materials.---}} 
Large phonon magnetic moments have been observed in Cd$_3$As$_2$~\cite{PhononMagnetization_20} and PbTe~\cite{PhononMagPbTe}. The former is a Dirac semimetal whereas the latter is a narrow gap semiconductor that is a close relative of SnTe, a topological crystalline insulator~\cite{TCI_SnTe_12}. 
We propose an effective model based on Cd$_3$As$_2$ in the presence of atomic displacement, which can also describe a trivial semiconductor at a different parameter.
The effective Hamiltonian reads~\cite{SM}
\begin{align}  \label{Hsm}
H_{\rm SM}(\bm{k})= \varepsilon_0+ \left[ \begin{array}{cccc} 
        \Delta & v_F \pi^\dagger &  \zeta \rho^\dagger  &         \\
       v_F \pi & -\Delta         &                      & -\zeta \rho^\dagger \\ 
  \zeta \rho   &                 &    - \Delta          & -v_F \pi^\dagger \\
               &  -\zeta \rho    & -v_F \pi             & \Delta 
             \end{array} \right]
\end{align}
where $\varepsilon_0(\bm{k})=\epsilon_0+\epsilon_1 k_z^2$ and $\Delta=\Delta_0 - \Delta_1 k_z^2$. In contrast to the model described by Eq.~\eqref{Heff}, the nontopological contribution here is three times larger than the topological one. We thus include both in following discussion.

For simplicity, we first study the case of $\Delta_0<0$ that corresponds to a semiconductor. 
The energy bands are plotted in Fig.~\ref{fig:Cd3As2V1}(a) with a band gap of about $40~$meV, which is much larger than the phonon energy in those experiments ($\sim 3~$meV). The adiabatic approximation is thus valid. The phonon induced magnetization from each $k_z$ is $\frac{eL_{\rm I}}{2m_{\rm I}}\frac{\zeta^2}{20\pi}\frac{-4}{\Delta^2}$ as plotted in Fig.~\ref{fig:Cd3As2V1}(b). 
By summing over these contributions and multiplying $NV_u$ ($N$ and $V_u$ are the number and volume of the unit cell in a sample respectively), the phonon magnetic moment can be obtained that is $\frac{eNL_{\rm I}}{2m_{\rm I}}\frac{\zeta^2}{80\pi}\frac{4}{\Delta_0 \sqrt{|\Delta_0\Delta_1|}}V_u$. By taking $\zeta\simeq 10~$eV/\AA$~$and $V_u \simeq 200~$\AA$^3$, this phonon magnetic moment can reach $10^4$ times of the atomic magneton ($ \frac{e\hbar}{2m_{\rm I}}\sim \frac{eNL_{\rm I}}{2m_{\rm I}}$). 

We then turn to the semimetal case with band inversion by setting $\Delta_0>0$. Two Dirac points appear where $\Delta(k_z)=0$ as denoted by the dashed lines in Fig.~\ref{fig:Cd3As2V1}(c). The magnetization from different $k_z$ is plotted in Fig.~\ref{fig:Cd3As2V1}(d), which increases in the manner of $\frac{\rm{sign}(\Delta)}{\Delta^2}$ as $k_z$ approach the Dirac points. Such divergence is due to the breakdown of the adiabatic approximation. Nevertheless, for such $k_z$ that $2|\Delta(k_z)|>E_p$ with $E_p$ being the phonon energy, the adiabatic approximation is still valid. By considering $k_z$ that satisfies the energy cutoff condition $2|\Delta| > E_p$, one can find that the phonon magnetic moment is larger than the above case by a factor of about $4{\rm log}\frac{4\Delta_0}{E_p}$ with a sign change.
By taking $E_p=3~$meV and $\zeta \simeq 1\sim 10~$eV/\AA, the magnetic moment is about $2\times10^{3\sim5}$ times larger than the atomic magneton $\frac{e\hbar}{2m_{\rm I}}$. 

One can generalize the model to describe three-dimensional strong and weak topological insulators~\cite{SM}. In these systems, one can also find a large phonon magnetic moment, which experiences a sign change when a strong topological insulator changes to a weak one.
}

\textit{\textbf{Summary.---}}
We have studied the phonon magnetic moment from the electronic orbital magnetization. 
We identified a topological contribution as a gauge-invariant second Chern form, which calls for the concept of a momentum-resolved Born effective charge and also contains a term from the phonon modified electronic energy coupled to the momentum space Berry curvature.
{For the chiral phonon in gaped graphene model, the topological contribution is the only source of the phonon magnetic moment, which can be large as the second Chern form corresponds to the gauge field near a Yang's monopole in this model. We also study the magnetic moment of optical phonons in bulk materials. We find large phonon magnetic moments in semimetal and narrow gap insulators, including weak and strong topological insulator. The orders of the phonon magnetic moments agree with recent experiments. In these systems, both topological and non-topological contributions are important.
}

\textit{Acknowledgements.---}
This work was supported by DOE (DE-FG03-02ER45958, Division of Materials Science and Engineering). Y.F. would like to thank the helpful discussion with Di Xiao, Shengying Yue, Haonan Wang, and Kaifa Luo.


\begin{thebibliography}{99} 

\bibitem{BornHuang_Book} M. Born and K. Huang, Dynamical Theory of Crystal Lattices (Oxford University Press, Amen House, London, 1962).

\bibitem{Phonon_AngularMom_Lifa_14} L. Zhang and Q. Niu, Angular Momentum of Phonons and the Einstein-de Haas Effect,
Phys. Rev. Lett. \textbf{112}, 085503 (2014).

\bibitem{Phonon_Chiral_AngularMom_Lifa_15} L. Zhang and Q. Niu, Chiral Phonons at High-Symmetry Points in Monolayer Hexagonal Lattices, Phys. Rev. Lett. \textbf{115}, 115502 (2015).

\bibitem{ChiralPhonon_Kagome_19} H. Chen, W. Wu, S. A. Yang, X. Li, and L. Zhang, Chiral phonons in kagome lattices, Phys. Rev. B \textbf{100}, 094303 (2019).

\bibitem{ChiralPhonon_Kekule_17} 
Y. Liu, C.-S. Lian, Y. Li, Y. Xu, and W. Duan, Pseudospins and Topological Effects of Phonons in a Kekul\'e Lattice, Phys. Rev. Lett. \textbf{119}, 255901 (2017). 

\bibitem{ChiralPhonon_r3xr3_18} 
X. Xu, H. Chen, and L. Zhang, Nondegenerate chiral phonons in the Brillouin-zone center of 
$\sqrt{3}\times\sqrt{3}$ honeycomb superlattices, Phys. Rev. B \textbf{98}, 134304 (2018).

\bibitem{ChiralPhonon_GhBN_18}  %
M. Gao, W. Zhang, and L. Zhang, Nondegenerate Chiral Phonons in Graphene/Hexagonal Boron Nitride Heterostructure from First-Principles Calculations, Nano Lett. \textbf{18}, 4424-4430 (2018).

\bibitem{Kagome_ChiralPhonon_21} A. Ptok, A. Kobia{\l}ka, M. Sternik, J. {\L}a$\dot{\rm z}$ewski, P. T. Jochym, A. M. Ole\'s, S. Stankov, and P. Piekarz, Chiral phonons in honeycomb sublattice of layered CoSn-like compounds, arXiv:2106.03740.

\bibitem{DanPhonon} D. Saparov, Y. Ren, B. Xiong, Q. Niu, Effect of Berry curvature on the dynamics of lattice, To be submitted.

\bibitem{ChiralPhonon_Exp_18} H. Zhu, J. Yi, M. Li, J. Xiao, L. Zhang, C. Yang, R. A. Kaindl, L. Li, Y. Wang, and X. Zhang, Observation of chiral phonons, Science \textbf{359}, 579 (2018).

\bibitem{ChiralPhonon_ExcitonReplicaTMD_19}  
Z. Li, T. Wang, C. Jin, Z. Lu, Z. Lian, Y. Meng, M. Blei, S. Gao, T. Taniguchi, K. Watanabe, T. Ren, S. Tongay, L. Yang, D. Smirnov, T. Cao, and S.-F. Shi, Emerging photoluminescence from the dark-exciton phonon replica in monolayer WSe$_2$, Nat. Commun. \textbf{10}, 2469 (2019). 

\bibitem{Phonon-Magnon_Exp_18} J. Holanda, D. S. Maior, A. Azevedo, and S. M. Rezende, Detecting the phonon spin in magnon-phonon conversion experiments, Nature Physics \textbf{14}, 500-506 (2018).

\bibitem{Phonon_MagneticField_17} T. F. Nova, A. Cartella, A. Cantaluppi, M. Först, D. Bossini, R. V. Mikhaylovskiy, A. V. Kimel, R. Merlin, A. Cavalleri, An effective magnetic field from optically driven phonons, Nat. Phys. \textbf{13}, 132-137 (2017).

\bibitem{Phonon_surface_21} R. Sasaki, Y. Nii, and Y. Onose, Magnetization control by angular momentum transfer from surface acoustic wave to ferromagnetic spin moments, Nat. Commun. \textbf{12}, 2599 (2021). 

\bibitem{PHE_Exp_05} C. Strohm, G. L. J. A. Rikken, and P. Wyder,  Phenomenological Evidence for the Phonon Hall Effect, Phys. Rev. Lett. \textbf{95}, 155901 (2005).

\bibitem{PHE_Exp_07} A. V. Inyushkin and A. N. Taldenkov, On the phonon Hall effect in a paramagnetic dielectric,
JETP Lett. \textbf{86}, 379 (2007).

\bibitem{PHE_SpinLiquid_Exp_17} K. Sugii, M. Shimozawa, D. Watanabe, Y. Suzuki, M. Halim,
M. Kimata, Y. Matsumoto, S. Nakatsuji, and M. Yamashita, Thermal Hall Effect in a Phonon-Glass Ba$_3$CuSb$_2$O$_9$, Phys. Rev. Lett. \textbf{118}, 145902 (2017).

\bibitem{THE_Cuprate_20} G. Grissonnanche, S. Th\'eriault, A. Gourgout, M.-E. Boulanger, E. Lefran\c{c}ois, A. Ataei, F. Lalibert\'e, M. Dion, J.-S. Zhou, S. Pyon, T. Takayama, H. Takagi, N. Doiron-Leyraud, and L. Taillefer,  Chiral phonons in the pseudogap phase of cuprates, Nat. Phys. \textbf{16}, 1108-1111 (2020).


\bibitem{PhononMagnetization_20} B. Cheng, T. Schumann, Y. Wang, X. Zhang, D. Barbalas, S. Stemmer, and N. P. Armitage, A Large Effective Phonon Magnetic Moment in a Dirac Semimetal, Nano Lett. \textbf{20} 5991 (2020).

\bibitem{PhononMagPbTe} A. Baydin, F. G. G. Hernandez, M. Rodriguez-Vega, A. K. Okazaki, F. Tay, G. Timothy Noe II, I. Katayama, J. Takeda, H. Nojiri, P. H. O. Rappl, E. Abramof, G. A. Fiete, and J. Kono, Magnetic Control of Soft Chiral Phonons in PbTe, arXiv:2107.07616. 

\bibitem{Phonon_OrbitMoment_19} 
D. M. Juraschek and N. A. Spaldin, Orbital magnetic moments of phonons, Phys. Rev. Materials \textbf{3}, 064405 (2019).

\bibitem{PhononBfield_Multiferroicity_17} 
D. M. Juraschek, M. Fechner, A. V. Balatsky, and N. A. Spaldin, Dynamical multiferroicity, Phys. Rev. Materials \textbf{1}, 014401 (2017).

\bibitem{PhononBfield_4fPM_20} D. M. Juraschek and P. Narang, Giant phonon-induced effective magnetic fields in $4f$ paramagnets, arXiv:2007.10556.

\bibitem{Cong_20} C. Xiao, Y. Ren, and B. Xiong, Adiabatically Induced Orbital Magnetization, arXiv:2012.08750.

\bibitem{Phonon_Spin_20} M. Hamada and S. Murakami, Conversion between electron spin and microscopic atomic rotation,
Phys. Rev. Research \textbf{2}, 023275 (2020).

\bibitem{Mag_Inhomogeneity_Di_09} D. Xiao, J. Shi, Dennis P. Clougherty, and Q. Niu, Polarization and Adiabatic Pumping in Inhomogeneous Crystals, Phys. Rev. Lett. \textbf{102}, 087602 (2009).

\bibitem{Dong_Niu_18} L. Dong and Q. Niu, Geometrodynamics of electrons in a crystal under position and time-dependent deformation, Phys. Rev. B \textbf{98}, 115162 (2018).

\bibitem{PhononMag_Vanderbilt_18} M. Stengel and D. Vanderbilt, Quantum theory of mechanical deformations, Phys. Rev. B \textbf{98}, 125133 (2018).

\bibitem{OrbMag_Adiabatic_19} L. Trifunovic, S. Ono, and H. Watanabe, Geometric orbital magnetization in adiabatic processes, Phys. Rev. B \textbf{100}, 054408 (2019).

\bibitem{NoteSpinOrb} See Eq.~(24) and Eq.~(11) in Ref.~\cite{Cong_20} and Eq.~(48) and Eq.~(50) in Ref.~\cite{Phonon_Spin_20}.

\bibitem{SM} In supplemental material, we include the details of the proof of the topological magnetization, the Wannier function contribution to the phonon magnetization, connection with previous theory, the effective Hamiltonian for the Dirac semimetal and topological insulators with phonon. References~\cite{Cd3As2SoftPhonon_19, Cd3As2Heff_17, Cd3As2Heff_13, TIHeff_09} are cited. 

\bibitem{BornEffCharge_DFT_97} X. Gonze and C. Lee, Dynamical matrices, Born effective charges, dielectric permittivity tensors, and interatomic force constants from density-functional perturbation theory, Phys. Rev. B \textbf{55}, 10355 (1997).

\bibitem{BornEffCharge_BerryCurv_02}
E. J. Mele and P. Kr\'al, Electric Polarization of Heteropolar Nanotubes as a Geometric Phase, Phys. Rev. Lett. \textbf{88}, 056803 (2002).

\bibitem{Phonon_ElectronCoupling_BornCharge_19} O. Bistoni, P. Barone, E. Cappelluti, L. Benfatto, F. Mauri, Giant Effective charges and Piezoelectricity in Gapped Graphene, 2D Mater. \textbf{6}, 045015 (2019).

\bibitem{Phonon_Polarization_20} D. Shin, S. A. Sato, H. H\"ubener, U. De Giovannini, N. Park, and A. Rubio, Nonlinear phononics in 2D SnTe: a ferroelectric material with phonon dynamical amplification of electric polarization, arXiv:2010.13646.

\bibitem{ModernTheory_Di_05} D. Xiao, J. Shi, and Q. Niu, Berry Phase Correction to Electron Density of States in Solids,
Phys. Rev. Lett. \textbf{95}, 137204 (2005).

\bibitem{ModernTheory_Resta_05} T. Thonhauser, D. Ceresoli, D. Vanderbilt, and R. Resta, Orbital Magnetization in Periodic Insulators, Phys. Rev. Lett. \textbf{95}, 137205 (2005).

\bibitem{ModernTheory_10} R. Resta, Electrical polarization and orbital magnetization: the modern theories, J. Phys.: Condens. Matter \textbf{22}, 123201 (2010).

\bibitem{Xiao_RMP} D. Xiao, M.-C. Chang, and Q. Niu, Berry phase effects on electronic properties, Rev. Mod. Phys. \textbf{82}, 1959 (2010). 
 
\bibitem{DiXiao_20} Yiqiang Zhao, Yang Gao, Di Xiao, Electric Polarization in Inhomogeneous Crystals, arXiv:2009.09306.

\bibitem{Note} The topological and non-topological contributions to the orbital magnetization can be obtained by expanding the Eqs.~(51) and (52) in Ref.~\cite{OrbMag_Adiabatic_19}. See supplemental materials for details.

\bibitem{OME_10} A. M. Essin, A. M. Turner, J. E. Moore, and D. Vanderbilt, Orbital magnetoelectric coupling in band insulators, Phys. Rev. B \textbf{81}, 205104 (2010).

\bibitem{YangMonopole_18} S. Sugawa, F. Salces-Carcoba, A. R. Perry, Y. Yue, and I. B. Spielman, 
Second Chern number of a quantum-simulated non-Abelian Yang monopole, Science \textbf{360}, 1429-1434 (2018).

\bibitem{SemimetalBC_16} H. Li, H. He, H.-Z. Lu, H. Zhang, H. Liu, R. Ma, Z. Fan, S.-Q. Shen, and J. Wang, Negative magnetoresistance in Dirac semimetal Cd$_3$As$_2$, Nat. Commun. \textbf{7}, 10301 (2016).

\bibitem{RenPRB15} Y. Ren, X. Deng, Z. Qiao, C. Li, J. Jung, C. Zeng, Z. Zhang, and Q. Niu, Single-valley engineering in graphene superlattices, Phys. Rev. B \textbf{91}, 245415 (2015).

\bibitem{ZhouShuyun1} C. Bao, H. Zhang, T. Zhang, X. Wu, L. Luo, S. Zhou, Q. Li, Y. Hou, W. Yao, L. Liu, P. Yu, J. Li, W. Duan, H. Yao, Y. Wang, and S. Zhou, Experimental Evidence of Chiral Symmetry Breaking in Kekul\'e-Ordered Graphene, Phys. Rev. Lett. \textbf{126}, 206804 (2021).

\bibitem{Graphene_Strain_Model_09} R. M. Ribeiro, Vitor M. Pereira, N. M. R. Peres, P. R. Briddon, and A. H. Castro Neto, Strained graphene: tight-binding and density functional calculations, New J. Phys. \textbf{11}, 115002 (2009).

\bibitem{TCI_SnTe_12} T. H. Hsieh, H. Lin, J. Liu, W. Duan, A. Bansil, and L. Fu, Topological crystalline insulators in the SnTe material class, Nat. Commun. \textbf{3}, 982 (2012).

\bibitem{Cd3As2SoftPhonon_19} S. Yue, H. T. Chorsi, M. Goyal, T. Schumann, R. Yang, T. Xu, B. Deng, S. Stemmer, J. A. Schuller, and B. Liao, Soft phonons and ultralow lattice thermal conductivity in the Dirac semimetal Cd$_3$As$_2$, Phys. Rev. Research \textbf{1}, 033101 (2019).


\bibitem{Cd3As2Heff_17} J. Cano, B. Bradlyn, Z. Wang, M. Hirschberger, N. P. Ong, and B. A. Bernevig, Chiral anomaly factory: Creating Weyl fermions with a magnetic field, Phys. Rev. B \textbf{95}, 161306(R) (2017).

\bibitem{Cd3As2Heff_13} Z. Wang, H. Weng, Q. Wu, X. Dai, and Z. Fang, Three-dimensional Dirac semimetal and quantum transport in Cd$_3$As$_2$, Phys. Rev. B \textbf{88}, 125427 (2013).

\bibitem{TIHeff_09} H. Zhang, C.-X. Liu, X.-L. Qi, X. Dai, Z. Fang, and S.-C. Zhang, Topological insulators in Bi$_2$Se$_3$, Nat. Phys. \textbf{5}, 438 (2009).

\end{thebibliography}
\end{document}